\documentclass[twocolumn]{aastex62}

\def\be{\begin{equation}}
\def\ee{\end{equation}}
\def\ba{\begin{eqnarray}}
\def\ea{\end{eqnarray}}

\def\12{{1\over 2}}

\def\msun{M_\odot}

\def\etal{{\it et~al.~}}
\def\ltsima{$\; \buildrel < \over \sim \;$}
\def\simlt{\lower.5ex\hbox{\ltsima}}
\def\gtsima{$\; \buildrel > \over \sim \;$}
\def\simgt{\lower.5ex\hbox{\gtsima}}

\received{...}
\revised{...}
\accepted{...}
\submitjournal{ApJ}

\shorttitle{Observing the influence by growing BHs}
\shortauthors{Vasiliev \etal}

\begin{document}

\title{OBSERVING THE INFLUENCE { OF GROWING BLACK HOLES ON THE PRE-REIONIZATION IGM} }

\correspondingauthor{Evgenii O. Vasiliev}
\email{eugstar@mail.ru}

\author{Evgenii O. Vasiliev}
\affiliation{Southern Federal University, Rostov on Don 344090, Russia}
\affiliation{Special Astrophysical Observatory, Russian Academy of Sciences, Nizhnii Arkhyz, Karachaevo-Cherkesskaya Republic 369167, Russia}

\author{Shiv K. Sethi}
\affiliation{Raman Research Institute, Sadashivanagar, Bengaluru 560080, Karnataka, India}

\author{Yuri A. Shchekinov}
\affiliation{Raman Research Institute, Sadashivanagar, Bengaluru 560080, Karnataka, India}
\affiliation{Lebedev Physical Institute, Russian Academy of Sciences, 53 Leninsky Ave., Moscow 119991}

\begin{abstract}
We consider cosmological implications of the formation of first stellar size black holes (BHs) in the universe. Such BHs form and grow by accretion in minihaloes of masses $\simeq10^5\hbox{--}10^7\msun$, and emit non-thermal radiation which impact the ionization and thermal state of the IGM. We compute the implications of this process. We show that the influence regions for hydrogen increase to 10kpc (physical length) for non-growing BHs to more than 0.3--1Mpc for accreting BHs, the influence regions are ten times smaller for singly ionized helium. We consider three possible observables from the influence zones around accreting BHs during $8.5<z<25$: HI 21cm line, hyperfine line of $^3$HeII, and HI recombination lines. We show that the 21cm emitting region around a growing BH could produce brightness temperatures $\simeq 15$mK across  an evolving structure of 1Mpc in size with hot, ionized gas closer to the BH and much cooler in outer regions. We show that the ongoing and upcoming radio interferometers such as LOFAR and SKA1-LOW might be able to detect these regions. $^3$HeII emission from regions surrounding the growing BH is weak: the corresponding brightness temperatures reaches tens of nano-Kelvin, which is below the range of  upcoming  SKA1-MED. We show that for growing BHs H$\alpha$ line could be detected by JWST with $S/N=10$ in $10^4$~seconds of integration. In light on the recent EDGES result, we show that with additional cooling of baryons owing to collision with dark matter the HI signal could be enhanced by more than an order of magnitude.
\end{abstract}


\section{Introduction} 

The probes of the epoch of reionization (EoR)  and cosmic dawn remain outstanding aims of modern cosmology. While relevant information about the era of cosmic dawn remains elusive,  important strides have been made in understanding the EoR since 2000, mainly owing to  the detection of Gunn-Peterson effect at $z \simeq 6$ and the CMB temperature and polarization anisotropies by WMAP and Planck \citep{Planck2015,Fanetal}. The discovery of Gunn-Peterson trough indicates  that the universe could be making a transition from  fully ionized to neutral  at $z\simeq 6$. The CMB anisotropy measurements are consistent with the universe being fully ionized at $z \simeq 8.5$. The current best bounds on the redshift of reionization from Planck put strong constraints on  the redshift of reionization,  $z_{\rm reion} = 8.5 \pm 1$ \citep{Planck2015}.

Theoretical estimates {show} that the first { stars} in the universe might have formed at { $z\simeq 65$} { \citep{naoz06}} thereby ending the  dark age of the universe. The emission of UV light from these structures carve out ionized regions which  might have percolated  at $z \simeq 9$ \citep[see e.g. ][and references therein]{Barkana2001}. However, the nature of these first sources that ionize and heat the intergalactic medium is difficult to establish within the framework of current theoretical models. The two mostly likely candidates are star-forming haloes and the precursors of quasars. In the latter case, the emission could be dominated by accretion onto  a seed  stellar-mass black hole, the case we consider in this paper. 

One  way to probe this phase is through  the detection of redshifted hyperfine transition of neutral hydrogen (\ion{H}{1}) from this era. The past one decade has seen  major progress on both theoretical and experimental efforts in this direction.  Theoretical estimates show that the global \ion{H}{1} signal is observable in both absorption and emission with its strength in the range $-200\hbox{--}20 \, \rm mK$ in a frequency range of $50\hbox{--}150 \, \rm MHz$, which corresponds roughly to a redshift range $25 > z > 8$ \citep[e.g. ][]{1997ApJ...475..429M,2000ApJ...528..597T,2004ApJ...608..611G,Sethi05,pritchard08,cohenfi}. The fluctuating component of the signal is likely to be an order of magnitude smaller on scales in the range $3\hbox{--}100$~Mpc { (comoving)}, which implies  angular scales in the range $\simeq 1\hbox{--}30$~arc-minutes \citep[e.g. ][]{Zaldarriaga2004}; \citep[for reviews see e.g. ][]{Zaroubi2013,Furlanettoetal,MoralesWyithe}. Many of the ongoing and upcoming experiments have the the capability to detect this signal in hundreds of hours of integration \citep[e.g. ][]{2015aska.confE...3A,2014MNRAS.439.3262M,parsons12,mcquinn06,morales05,kulkarni16,pen09}. Upper limits on the fluctuating component of the \ion{H}{1}  signal have  been obtained by many ongoing experiments --- GMRT, MWA, PAPER, and LOFAR \citep{2017ApJ...838...65P,2016ApJ...833..102B,PAPER,GMRT}. 

In addition to the redshifted hyperfine line of HI, it might be possible to probe cosmic dawn and EoR using other spectral lines of the primordial gas. Therefore, we consider also HI recombination lines and hyperfine line of $^3$HeII. 

In this paper, we consider the impact of a {growing} black hole (BH)on the thermal and ionization state of the IGM in the redshift range {$8 < z <25$}. {There is  copious observational evidences of the existence of supermassive black holes with masses  upto $M\sim 10^9~\msun$ at $z\simeq 7$  \citep[see e.g.,][]{mortlock11,banados,shellqs-survey,viking-survey,panstarrs-survey,wu-bhs15}\footnote{http://www.homepages.ucl.ac.uk/~ucapeib/list\_of\_all\_quasars.htm}. The presence of such ``monstrous'' black holes in the young Universe with  ages less than 500 Myr seems challenging because of strong radiative and wind feedback \citep[see in][]{bhgrowth-illustris,massiveblack-sim,bhgrowth-eagle,bhgrowth-horizonAGN,negri17,gaspari17,bhgrowth-illustristng,latif16,latif18}. In this paper we address the question of whether the regions around these growing  BHs can be observed  in 21~cm emission, helium hyperfine line and hydrogen recombination lines. }

In the next section, we describe our model of photon emission from a BH that forms in the redshift range $20\hbox{--}25$ and subsequently grows owing to accretion. In section~\ref{sec:obser} we discuss possible observables that can probe the thermal and ionization evolution of the gas influenced by emission from the BH. In section~\ref{sec:resu} we present our main  results.  In section~\ref{sec:sumcon} we summarize our findings and make concluding remarks. Throughout this paper, we assume the spatially-flat $\Lambda$CDM model with the following parameters: $\Omega_m = 0.254$, $\Omega_B = 0.049$, $h = 0.67$ and $n_s = 0.96$, with the overall normalization corresponding to  $\sigma_8 = 0.83$ \citep{Planck2015}.

\section{Description of the model}

The accretion onto a black hole (BH) is supposed to be  a source of UV/X-ray photons. {Supermassive black holes (SMBH) with masses $\simgt 10^9M_\odot$ are known to exist  at redshifts as high as $z>7$ \citep{mortlock11,banados}. One can expect that during their growth phase  their predecessor would contribute to heating and ionization of the Universe.  In order that a stellar-mass BH seed would grow to a $10^9M_\odot$ SMBH, a nearly continuous accretion with the Eddington rate is the most efficient regime, under which  the BH mass $M_{BH}$ grows as} \citep{shapiro05,volonteri05,volonteri06}:
\be \label{epsi}
 M_{BH}(t) = M_{BH,t=0} {\rm exp}\left({1-\epsilon \over \epsilon} {t \over t_{E}}\right)
\ee
where $M_{BH,t=0}$ is initial BH mass and  $t_{E}=0.45$~Gyr, $\epsilon$ is radiative efficiency{ ---the efficiency of conversion of rest-mass energy to luminous energy by accretion onto a black hole of mass $M$ \citep{shapiro05}}, $\epsilon \simeq  0.1$ is taken as fiducial value; we discuss the impact of varying $\epsilon$ in later sections. { Following \citet{shapiro05} we assume that the efficiency of accretion luminosity $\epsilon_L \equiv L/L_E$, where $L_E$ is the Eddington luminosity, is equal to unity in our calculations;} { $\epsilon_L=1$ is thought to represent the upper observed limit of quasar luminosity \citep{mclure04}. }

The spectrum of the ionizing radiation emitted during accretion is assumed  to be  a power-law:
\be
 L_\nu = L_0 \left({\nu \over \nu_{\rm H}}\right)^{\alpha}
 \label{lum}
\ee
where $\alpha = -1.5$,{ which is assumed as a fiducial value in our calculations}, $L_0$ is a normalization coefficient, which is obtained for the bolometric luminosity of BH: $L_{BH} = 1.25\times 10^{38} M_{BH}$~erg/s in energy range from 13.6 to $10^4$~eV. The bolometric luminosity is assumed to be equal to the Eddington limit. The spectral energy distribution slope of active galactic nuclei is measured to be from $-1.7$ to $-1.4$ \citep{telfer02,scott04,shull12,stevans14,lusso15}, and  we consider how this affects on our results below.
 
Current theoretical models of first (Population III) in the Universe favor the IMF to be dominated by massive objects in the range from tens to hundreds of solar masses \citep[e.g.,][]{abel02,bromm02,yoshida08}. In the lower mass end (tens of solar masses) stars can  form either due to various feedbacks \citep{tan04,hosokawa11}, or due to atomic cooling in metal-free gas with the virial temperature $T>10^4$~K \citep[e.g.,][]{becerra15}, or owing  to cooling by metals/dust in a weakly enriched gas \citep[e.g.,][]{bromm01,dopcke13}.  

The seeds for BHs are the final product of the evolution of Population III stars with $M \sim 30\hbox{--}260\msun$  \citep[see e.g.,][]{woosley02}. Low-mass stars, $M \sim 30\msun$, are likely more numerous and might be more common seeds for BHs. However, only a small mass fraction $\sim 10$\% of their progenitors collapses to a BH and therefore these star do not contribute significantly to the growth of  supermassive black holes. Even though this fraction increases for higher mass stars, it still remains less than 50\% for $M \simlt 100\msun$. However, stars with $M\simgt 260$~$\msun$ leave remnants---black holes, only slightly less massive than the progenitor. The Eddington rate is relatively slow: as seen from Eq. (\ref{epsi}) the accretion with $\epsilon=0.1$ increases the BH mass by factor 10 from $z=50$ to $z\simeq 20$ and by 30 to $z\simeq 17$. Therefore, only stellar progenitors of $M\simgt 260$~$\msun$ are capable  of giving  rise to SMBHs. Based on these considerations, in our calculations, we adopt $M_{BH}=300~\msun$ as fiducial value for BH seeds, though deviations from this value are also discussed. It is worth noting that in low-mass halos \citep[apparently $<10^6M_\odot$][]{wfryer12,jeon12,ricot11}, radiative and mechanical feedback can inhibit growing supermassive BH from a stellar mass BH seed. However, currently there are too few numrical simulations for firm conclusions about inhibitive feedbacks on the growth of SMBHs in more massive $\geq 10^8M_\odot$) minihalos  \citep{wise18}. 

Stellar progenitors of BHs are formed in minihalos with total masses $M\sim 10^5-10^7~\msun$ \citep{haiman96,tegmark97},  \citep[see also review][]{Barkana2001}. {Eventually, depending on specific conditions in a minihalo a single very massive star or/and several less massive stars do form and produce copious amount of ionizing photons \citep{tum00,brom04}. As a result a significant fraction of gas in the host minihalo becomes ionized, and the escape fraction of Lyman continuum photons into the IGM can grow substantially \citep[see review][]{ciardi05rev,ferrara13}}.  

To model absorption of ionizing photons inside the halo we assume that average total column density of HI inherent to the host galaxy is $N_{\rm HI}^{h}$ with primordial abundance of elements:  $X=0.76$, $Y_{\rm He}=0.24$. { In calculations we include not only absorption in the host  galaxy, but in the circumgalactic gas within several virial radii ($\simeq 1\hbox{--}3$)  as well. Neutral hydrogen fraction in the interstellar medium (ISM) of the host halo is determined by detailed evolution of the halo, e.g. gas coolng/heating, possible star formation and BH feedbacks, and so on. On the other hand, density and velocity profiles outside the halo might be altered  by tidal interactions and merging with other halos. Therefore,  finding a closer connection between $N_{\rm HI}^h$ and the underlying galactic ISM is very challenging \citep[see e.g.,][]{bromm03,whalen04,greif07,whalen08,v08,v12}, and lies out of the scope of the paper. However, it is obvious that on much larger scales where the diagnostics discussed in this paper arise, e.g.  21cm signal,  details of  gas distribution around host halo plays a minor role and only the average value of $N_{\rm HI}^h$ might suffice to model the absorption inside the halo.} 

Therefore, we consider several values for $N_{\rm HI}^h$ to model the  host halo, with $N_{\rm H}^h = 10^{20}$~cm$^{-2}$ as a fiducial value. This  choice is consistent with the fact that the total column density of minihalos with $M\simlt 10^9\msun$ formed at $z\simeq 10\hbox{--}20$ are less than $10^{21}$~cm$^{-2}$ (assuming the top-hat density profile, for simplicity). As we expect  the gas inside halos to be  partially ionized by both stellar progenitors and the BH itself, the adopted value of HI column density seems a reasonable conservative estimate. In Sec. \ref{ipar} we will discuss dependence of  our results on its variation.

The radiation from the growing BH can also be attenuated by neutral IGM gas. { Optical depth at distance $r$ from the BH is $\tau_\nu(r) = \int {\sigma_\nu^k(r) n^k(r) dr }$, where $k = {\rm HI, \ HeI, \ HeII}$, $\sigma_\nu^k$ are the cross-sections at frequency $\nu$ \citep{cen92,glover07}, the values $\sigma_\nu^k(r)$ and $n^k(r)$} depend on ionization and thermal history of the IGM, whose evolution is described below.

{ Then, t}he flux { of ionizing radiation} at a distance $r$ from the BH is 
\be
 F_\nu = {L_\nu \over 4 \pi r^2} {\rm exp}(-\tau_{\rm h}-\tau_{\rm IGM})
\ee
where the first term in the exponent is due to the attenuation in the host galaxy (it depends on $N_{\rm HI}^h$ and we assume it to remain constant during the evolution), the second term is determined by absorption in the medium surrounding the BH  and the  intergalactic medium.

In the hierarchical structure formation  scenario, minihalos undergo mergers. {In some of which the seeds of BHs with intermediate mass form, and can efficiently grow only when a considerable reservoir of gas is available. It suggests that minihalos with growing BHs undergo frequent mergers, and collect a sufficient gas mass for BHs feeding. The lower the radiative efficiency during a BH growth the larger is the mass necessary to maintain the growth, and it may happen that this mass will exceed the initial baryon mass of the host minihalo.} For instance, for $\epsilon=0.2$ a BH mass grows about 33 times in $\sim$400~Myr. For a seed with $M_{BH}=300~\msun$ such a growth can be maintained in the host minihalo {as small as $M\simlt  10^5~\msun$. However, for $\epsilon=0.1$ in the same period} the BH mass increases by about $2.5\times 10^3$ times, whereas for $\epsilon=0.05$ the ratio reaches values as high as $1.5\times 10^7$. {Therefore, only those minihalos with the accretion rate higher than $\dot M\geq 0.4~\msun$ yr$^{-1}$ could host growing BH with such a low radiative efficiency.}

In the $\Lambda$CDM model, star-forming minihalos for a wide range of masses $< 10^8 \, \rm \msun$ merge and virialize at $z\simgt 25$ as  (3-4)$\sigma$ density peaks. {The merging rate of such minihalos seems to be sufficient} to provide the sites for feeding growth of massive BHs \citep{volonteri05}. Based on these considerations  we start the evolution at $z_0=25$ and continue it for  400~Myr, which  corresponds to the final redshift $z\simeq10$, which is close to the era at which reionization of the Universe is completed. Therefore we explicitly assume that black holes grow nearly with a steady state Eddington accretion rate on cosmological time scales. In this regard such a consideration excludes a possibility to incorporate here recently widely discussed direct monolithic collapse of supermassive black holes \citep{begel06}, as they apparently keep the accretion rate close to the Eddington limit only in a very short time scale $\sim 1$ Myr after formation \citep[see e.g.][]{john11}.

How numerous could the high redshift BHs be? The space density of haloes that can host BHs at high redshifts can be computed using Press-Schechter formalism.  Assuming  a typical halo mass of $10^7 M_\odot$, the comoving  density of such haloes increases from $10^{-2} \, \rm \, Mpc^{-3}$ to nearly $1 \, \rm  Mpc^{-3}$ in the redshift range 10 to 20 \citep[e.g.][]{Barkana2001}. However, converting space density of haloes that can host BHs to the number density of BH precursors is highly uncertain.  Their  comoving density could lie in the range $10^{-3}\hbox{--}10^{-10} \, \rm Mpc^{-3}$ at $z \simeq 10$  \citep[e.g. Figure~4 of ][]{D14}.

Two AGNs have been  detected at $z>7$ and these AGNs host BHs with  $M\simeq 10^9 \, \rm  M_\odot$. In the  case  these BHs grew from stellar seeds, the growth could have commenced at
$z\simeq 14$ to reach $M \simeq 10^9 \, \rm M_\odot$ within one Eddington time (0.45~Gyrs, which is close our final time in calculations) for  $\epsilon = 0.05$. At smaller redshifts ($z\simeq 2\hbox{--}4$) such AGNs have absolute magnitudes in the range -26 to -28 (see Figure 5 in \citet{mclure04} and Figure 13 of \citet{qso-lumfunc}). Less massive BHs $M \simeq 10^8 M_\odot$ are expected to be more than hundred times more numerous \citep[Figure 6 in][]{volonteri06}. These BHs could have emerged from the same mass haloes but for larger values of $\epsilon$.

In our mode, a minihalo with a seed BH is immersed into the IGM. {Dynamical state and structure of the transition layer between minihalo and surrounding gas can be in general complicated, 
with inhomogeneous distribution of  gas density, temperature and velocity field.}  We neglect the complications of this narrow interface and {match the minihalo 
directly to the IGM, starting our calculations from the internal boundary of the surrounding intergalactic gas. We assume this  gas to have homogeneous distribution of density 
and temperature decreasing due to cosmological expansion:  $\propto (1+z)^{-3}$ and $\propto (1+z)^{-2}$,  {until} the ionizing radiation 
from a BH  changes its thermodynamics}. 

We consider the evolution of gas enclosed in the concentric static spheres with a BH in the center. The radii of spheres  extend from $10^3$ to $10^7$~pc { (all distances are expressed in physical units unless otherwise specified)}. The radii of neighbouring spheres differ by a factor $a_r=1.1$: $r_{i+1} = a_r r_i$, which  yields   100 concentric shells accounting  for the ratio of outer radius to inner  radius. Note that the inner radius is about three times greater than the virial radius of minihalo with $M=10^7~\msun$ formed at $z=20$ \citep[e.g.][]{ciardi05rev}.

In each sphere we solve thermal and ionization evolution of hydrogen, neutral helium and singly ionized helium. We consider the following processes for primordial plasma: collisional ionization, recombination, photoionization by UV/X-ray radiation from the BH attenuated by both the host galaxy and the surrounding IGM gas. The thermal evolution  includes cooling due to collisional ionization for HI, HeI, HeII, recombination of HII, HeII (radiative and dielectronic), HeIII, collisional excitation of HI, HeI ($1^2S$ and $2^3S$), HeII, free-free emission and Compton cooling/heating, and  photoionization heating. The reaction and cooling/heating rates are taken from \citet{cen92,glover07}. Because we consider ionization by X-ray radiation the influence by secondary electrons is taken into account as described in \citep{steenberg85,ricotti02}. In {equation of thermal evolution  we add the cooling term due to the Hubble expansion, in order to correctly describe evolution} on time scales greater than the local age of the universe. We solve the equations on time scale 400~Myr, such that for the initial redshift $z_0=20$ our calculations complete at $8.5$. The initial gas temperature and HII fraction for a given redshift are  obtained by using the RECFAST code \citep{recfast}, while helium in the initial state is assumed to be neutral.

\section{observable features of Cosmic Dawn and Epoch of Reionization} \label{sec:obser}
In this section we discuss in detail the possible observables in the redshift
range $8.5 < z < 25$ owing to the impact of radiation from the accreting BH. 

\subsection{21cm line}

Atomic collisions and scattering of UV photons couple the HI spin temperature { to} the gas kinetic temperature, $T_k$, { and the color temperature, $T_c$} \citep{field,wout}:
\be
 T_s^{\rm HI} = {T_{CMB} + y_c T_k + y_a T_c \over 1 + y_c + y_a}
\label{tspin}
\ee
Here $y_c$ and $y_a$ determine the coupling of { the} two states of the HI hyperfine splitting owing to collisions and Lyman-{ $\alpha$} photons (Wouthuysen-Field coupling), { respectively}; $y_c = C_{10}^{\rm HI} T_\star/(A_{\rm HI}T)$ with $T_\star = h\nu_{\rm HI}/k$ and $C_{10}^{\rm HI}$ being the collisional de-excitation rate of the hyperfine line of HI. { We assume that the color temperature is coupled to the gas kinetic temperature: $T_c \simeq T_k$.}

The coefficient $y_a$ is similarly defined  with collisional de-excitation rate replaced with the de-excitation rate owing to Lyman-$\alpha$ photons.  Given the geometry of our physical setting, the coupling of the expanding gas with Lyman-$\alpha$ photons from BH needs to be discussed in detail. Lyman-$\alpha$ photons in the rest frame of BH are strongly absorbed in the halo of column density $N_{\rm HI} \simeq 10^{20} \, \rm cm^{-2}$ surrounding the BH as the line center cross-section for Lyman-$\alpha$ scattering is $\simeq  10^{-13} \, \rm cm^2$ (assuming 
a temperature $T\simeq 5000 \, \rm K$). Using Voigt profile one can show that photons of frequencies $\nu \simeq \nu_\alpha\pm 50\Delta\nu_D$, with $\Delta\nu_D=\nu_\alpha/c(2kT/m_p)^{1/2} \simeq 10^{-5} \nu_\alpha$ being the Doppler width -- can escape the halo. As the medium outside the halo is expanding the photons will redshift, and photons with frequencies larger than Lyman-$\alpha$ in BH rest frame  can get absorbed in the expanding medium. Using local Hubble's law, $v = H(z)r$ one can show these photons to get absorbed for a range of distances $0.01\hbox{--}1 \, \rm Mpc$ from the halo. 

This motivates us to assume that the number of Lyman-$\alpha$ photons (which are photons with frequencies marginally above Lyman-$\alpha$ in BH rest frame) in the expanding medium suffer only
geometric $1/r^2$ dilution. In addition, we also assume that {\it in situ} injected Lyman-$\alpha$ photons emerge due to recombinations with number density proportional to the local photoinization rate \citep[see~Eqs.~15~and~17~in][]{miralda04}. Here following \citet{field}  we also explicitly assume the ``color temperature'' of Ly-$\alpha$ photons to be equal to the gas kinetic temperature. 

In the collisional de-excitation rate we take into account collisions with H atoms \citep{kuhlen} and electrons \citep{liszt}.  $y_a$ is proportional to the number density of Lyman-$\alpha$ photons at the point of scattering.

The differential brightness temperature  for the redshifted HI line  can be estimated \citep{miralda04,Furlanettoetal}: 
\ba
   \Delta T^b_{\rm HI} = 25~{\rm mK} (1 + \delta)  {n_{\rm HI} \over n} ~  {T_s^{\rm HI} - T_{CMB} \over T_s^{\rm HI}}
                      \left({\Omega_b h \over 0.03}\right)   \times \nonumber \\
               \times      \left({0.3 \over \Omega_m}\right)^{0.5}  \left({1+z\over 10}\right)^{0.5}                 
                     \left[ { H(z)/(1+z) \over dv_{||}/dr_{||} }\right]
\label{tb21}
\ea
where $\delta$ is overdensity, which is neglected as we assume { the} uniform Hubble expansion at high redshifts so that the gradient of the proper velocity along the line of sight $dv_{||}/dr_{||}$ equals  $H(z)/(1+z)$. { Near the halos hosted by BHs the line broadening is dominated by peculiar velocities of IGM gas rather than the Hubble expansion, in this case the center-of-line optical depth and consequently the brightness temperature $\Delta T^b_{\rm HI}$ decrease. }

\subsubsection{Global condition on HI absorption  from EDGES observation}

In this work, we model the HI signal from gas surrounding an isolated  accreting black hole. Such black holes are not the only source that can emit UV radiation relevant for modelling HI absorption and emission signal from high redshift. While we do not incorporate in our models all other possible sources in the redshift range of interest, it is important to know physical state of gas far from the zone of influence of the black hole. This would allow us to smoothly match HI signal from regions close to the black holes to the signal expected from  ambient gas under global conditions. 

Recent EDGES observation \citep{2018Natur.555...67B} shows a sky-averaged absorption feature of strength $\Delta T \simeq -500 \,\rm mK$ in the frequency range $70$--$90$~MHz which 
corresponds to a redshift range $15$--$19$ for the redshifted HI line. 

The minimum temperature of the IGM at $z\simeq 19$ is $T \simeq 6 \, \rm K$ in the usual case (standard recombination history), and it follows from Eq.~(\ref{tb21}) that the absorption trough in the redshifted HI hyperfine line at $z\simeq 19$  should not have been deeper than around $-180 \, \rm mK$. One  plausible explanation of EDGES results  relates to overcooling of baryons by elastic collisions with dark matter particles, as suggested by \citet{2018Natur.555...71B}. In this case, as seen in Eqs.~(\ref{tspin}) and~(\ref{tb21}):  ({\it a}) Lyman-$\alpha$ photons globally couple the spin temperature to matter temperature, i.e. $T y_\alpha \gg T_{CMB}$, such that $T_s = T$ at $z\simeq 19$, and ({\it b}) $T_s \ll T_{CMB}$ as the signal  is seen in absorption and is strong. Note though that this explanation is still widely debated, because of a possible systematic error of the EDGES result  \citep{hills18}. Another possible  alternative explanations might be  that there is additional  radio background at $z\simeq 18$ whose temperature $T_{\rm radio}$ is higher than the CMB temperature; in this case we can replace $T_{\rm CMB}$ with the $T_{\rm CMB} +T_{\rm radio}$ in Eq.~(\ref{tb21}) \citep{feng18} and the enhancement of the observed signal is not owing to the cooling of baryons. It is also conceivable that  the observed feature is owing to  radiation from spinning dust grains in the Galactic ISM  \citep{draine-miralda}.  We explore here the implications of coupling between DM and baryons as a possible explanation of EDGES result.

To model the global conditions implied by EDGES observations we  solve one more equation for the dark matter  temperature, $T_{\rm dm}$, which can be altered due to adiabatic expansion and the interaction with baryons. In addition,  the term corresponding  to the interaction of matter with dark matter is added to the matter temperature equation as well \citep[for details see e.g.][]{2018arXiv180303091B}. We follow \citet{2018arXiv180303091B} in modelling $\sigma_{\rm dm}$, the energy-exchange cross-section between dark matter and matter,
as following the form of Rutherford scattering between a millicharge dark matter particle with electrons \citep[e.g.][]{dm-charged1,dm-charged2,dm-charged3,dm-charged4,dm-charged5}. In this case, $\sigma_{\rm dm} = 8\pi g^2 e^4/(m_e m_{\rm dm} v^4) \log{(\Lambda)}$, here $g$ is { the} ratio of dark matter charge to electron charge. $v$ is the relative velocity between the two particles and $\langle \sigma_{\rm dm} v \rangle$ corresponds to thermal averaging. The most significant aspect of such an interaction for our purposes is that  it is proportional to  $1/v^4$ and therefore at higher redshifts when the temperature is higher, the interaction is negligible. $n_{\rm dm}$ the number density of dark matter particles is another free parameter, $n_{\rm dm} = \rho_{\rm dm}/m_{\rm dm}$. 

As there are three free parameters in modelling this interaction---the number density of dark matter particle (or equivalently the mass of dark matter particle) and interaction strength  between dark matter and baryons, and initial temperature of dark matter---a rich array of possible scenarios are possible  \citep[for details see e.g.][]{2018arXiv180303091B}. It is not our aim here to constrain these parameters but obtain the global condition of the HI gas in the redshift range of interest. { To  cool the baryons at $z\simeq 20$, we require, $n_{\rm dm} \sigma_{\rm dm} v > H$, where $H$ is the expansion rate of the universe. Using the expression for $\sigma_{\rm dm}$ given above, $v$ as  the average speed  of thermal electrons at $z\simeq 20$ before the additional cooling sets in, and assuming  $1\%$ of the DM to be milli-charged, we obtain $g>10^{-7}$
for  $m_{\rm dm} \simeq 10 \, \rm MeV$, in agreement with the results of \citet{2018arXiv180303091B}.}

We also need a global  heating source which allows the HI to heat above the CMB temperature for $z<15$, as the EDGES observation require. We add the corresponding term  that gives additional  heating  source for  global heating; this term is modelled as photoelectric heating by x-ray photons from  sources  except from the BH  (see e.g. \citep{barkanafi,cohenfi,fialkov18}. 

As noted above,  an essential component of our modelling the EDGES result  is $T_s^{\rm HI}=T$ in Eq. (\ref{tspin}). All the parameters (global heating rate, cross-section of dark matter-baryon scattering, Lyman-$\alpha$ coupling)  have  been chosen  to fit the 21 cm brightness temperature $T_b \sim -500$ mK at $z\sim 20$ as in \citep{2018Natur.555...67B}.

\subsection{$^3${\rm HeII} hyperfine line}
Another important hyperfine structure transition exists in a singly ionized helium-3 isotope $^3$HeII at 8.67~GHz\citep{townes57,sunyaev66,goldwire67,rood79,bell00,bagla09,mcquinn09,takeuchi14}. Similar to the HI hyperfine line this transition is excited by collisions with atoms and electrons and  photon scattering. The rate of transition owing  to collisions  for a singly ionized helium-3  with electrons is given by:  
\be
 C_{10}^{\rm ^3HeII} = n_e \left( {k T \over \pi m_e c^2} \right)^{1/2} c \sigma_e^{{\rm ^3HeII}}
\ee
where $\sigma_e^{{\rm ^3HeII}}$ is the average cross-section of the spin exchange between $^3$HeII and electrons, which is approximated as (McQuinn \& Switzer 2009)
\be
 \sigma_e^{ {\rm ^3HeII}} \simeq {14.3 {\rm eV} \over k T} a_0^2,
\ee
where $a_0$ is the Bohr radius. In this case, the Wouthuysen-Field coupling between the two levels is caused by photons of wavelength, $\lambda = 304 \, \rm \AA$ \citep[Eq.~17 in][]{mcquinn09}. 
The number density of these photons at the point of scattering is computed from
the spectrum of BH emission.  This allows us to calculate  the differential brightness temperature of $^3$HeII  line  using  Eq.~(\ref{tspin}):
\ba
   \Delta T^b_{^3{\rm HeII}} = 1.7\times 10^{-3}~{\rm mK} ~ (1 + \delta) ~  {n_{\rm HeII} \over n} \times \nonumber \\
                             \times  ~  {T_s^{\rm ^3HeII} - T_{CMB} \over T_s^{\rm ^3HeII}} \left( {Y_{\rm ^3He}\over 10^{-5}}\right)
                      \left({\Omega_b h \over 0.03}\right)   \times \nonumber \\
               \times      \left({0.3 \over \Omega_m}\right)^{0.5}  \left({1+z\over 10}\right)^{0.5}                 
                     \left[ { H(z)/(1+z) \over dv_{||}/dr_{||} }\right]
\label{tbhe}
\ea
where $Y_{\rm ^3He}$ is the primordial abundance of the helium-3 isotope, which is assumed to be equal $10^{-5}$, $n_{\rm HeII}$ is the number density of singly ionized helium-4 isotope.

\subsection{Optical and radio- recombination lines}

As seen in Figure~\ref{fig-evol}, growing BHs produce regions of high ionization which can potentially be detected in hydrogen recombination lines. 

The frequencies of H$nj$ lines are: $ \nu_{{\rm H}nj} = c R \left[ {1\over n^2} - {1\over j^2} \right]$ with $j>n$ and $R = 1.0968\times 10^5$~cm$^{-1}$ being the Rydberg constant for hydrogen.  The emissivity of a recombination line averaged over   a sphere of radius $r_i$ around the BH is: $\epsilon_{{\rm H}nj}(r_i) = q_{nj} \alpha_j  n_e(r_i) n_{\rm HII}(r_i)$;  $n_e(r_i)$ and $n_{\rm HII}(r_i)$ are number densities of $e$ and HII species, respectively, $\alpha_j$ is the recombination coefficient to $j$th state \citep[for detailed discussion and derivation see e.g.][]{rybicki-book}, and $q_{nj} \simeq A_{jn}/\sum\limits_{m<j}^{} A_{jm}$ is the probability that an atom recombined to  $j$th state emits a photon by a spontaneous decay on to $n$th state. {In practice ${\rm H}n\alpha$ lines from transitions between the states $n$ and $n+1$ are usually considered because they are the strongest with the $A$-coefficients being the largest \citep[e.g.][]{rybicki-book}. }

The emissivity in these lines can be approximated as: $\epsilon(r_i) \simeq 3.25 n^{-2.72} \alpha_B(r_i) n_e(r_i)  n_{\rm HII}(r_i)$ \citep{rule13}, where $\alpha_B$ is the case B recombination rate \citep[eqn. 14.8 in][]{drainebook}.  The total luminosity in $j$-line is $L_j = (h c / \lambda_j) \sum_i  \epsilon_j(r_i) V(r_i)$, where $V_i$ is the volume of $i$-sphere. We include all spheres with $T>100$~K and achieve reasonable convergence for the predicted luminosity {as the ionized fraction falls faster than $r_i^{-2}$}.  The flux in $j$-line is 
\be
F_{\nu_j} = {1 \over \Delta\nu_j} ~ { (1+z) L_j \over 4 \pi d_L^2}
\label{flux}
\ee
where $d_L$ is the luminosity distance. {The line width is $\Delta\nu_j = \max(1/\sum_{i<j} A_{ni},\Delta\nu_D)$.}  {In all cases of interest here the Doppler broadening dominates, such that $\Delta\nu_j$ is}  given by Doppler line width:
\be
 \Delta\nu_j = {\nu_{j0} \over c} \sqrt{2 k T \over m_p}.
\label{width}
\ee

\section{Results} \label{sec:resu}

\subsection{{\rm HI} 21 cm observables} \label{sib:21cm}

\begin{figure}
\center
\includegraphics[width=75mm]{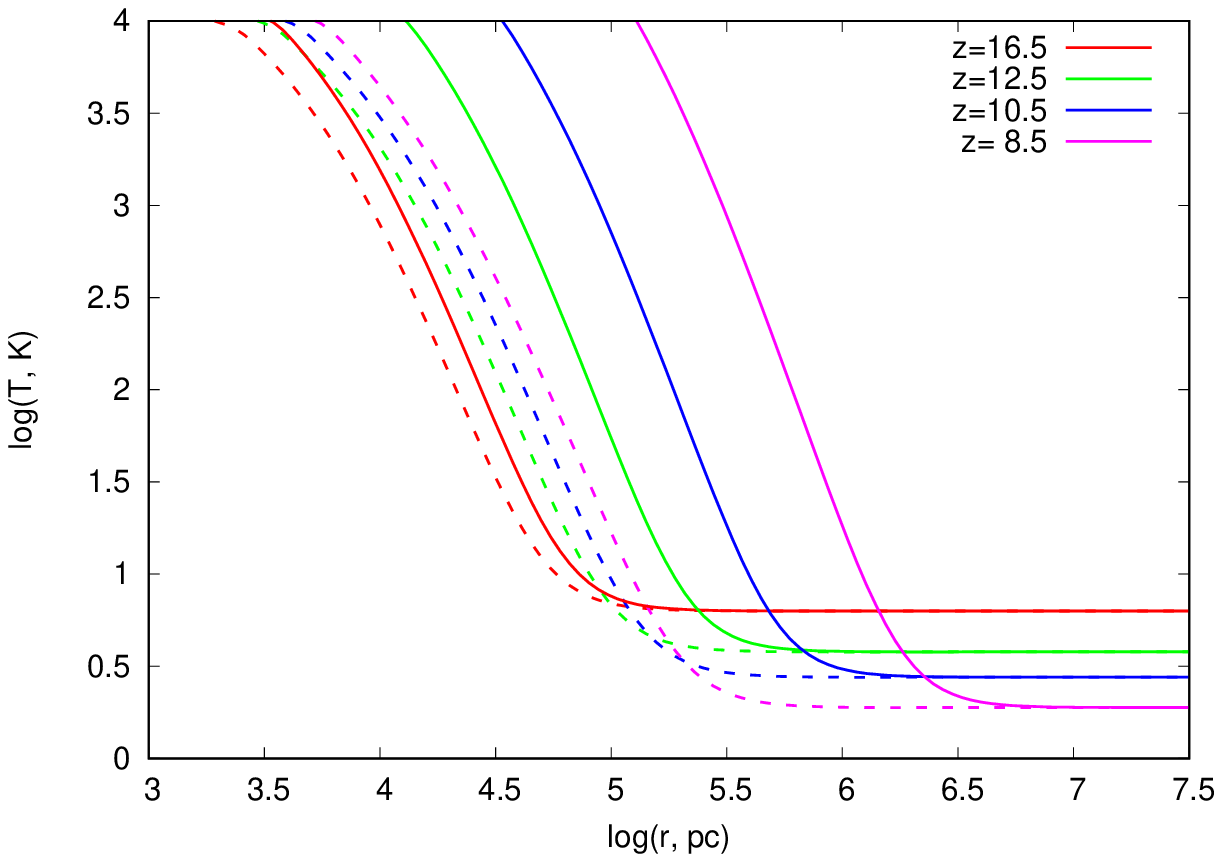}
\break
\includegraphics[width=75mm]{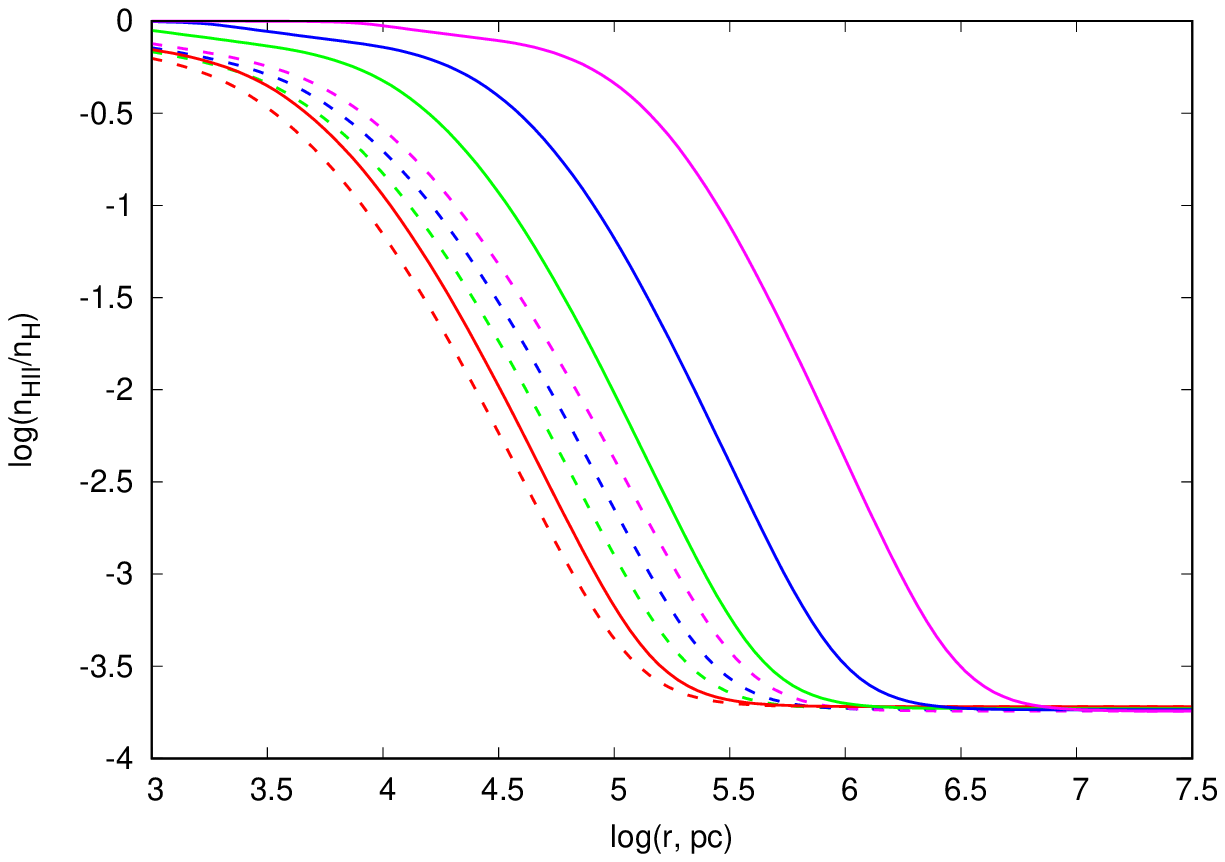}
\break
\includegraphics[width=75mm]{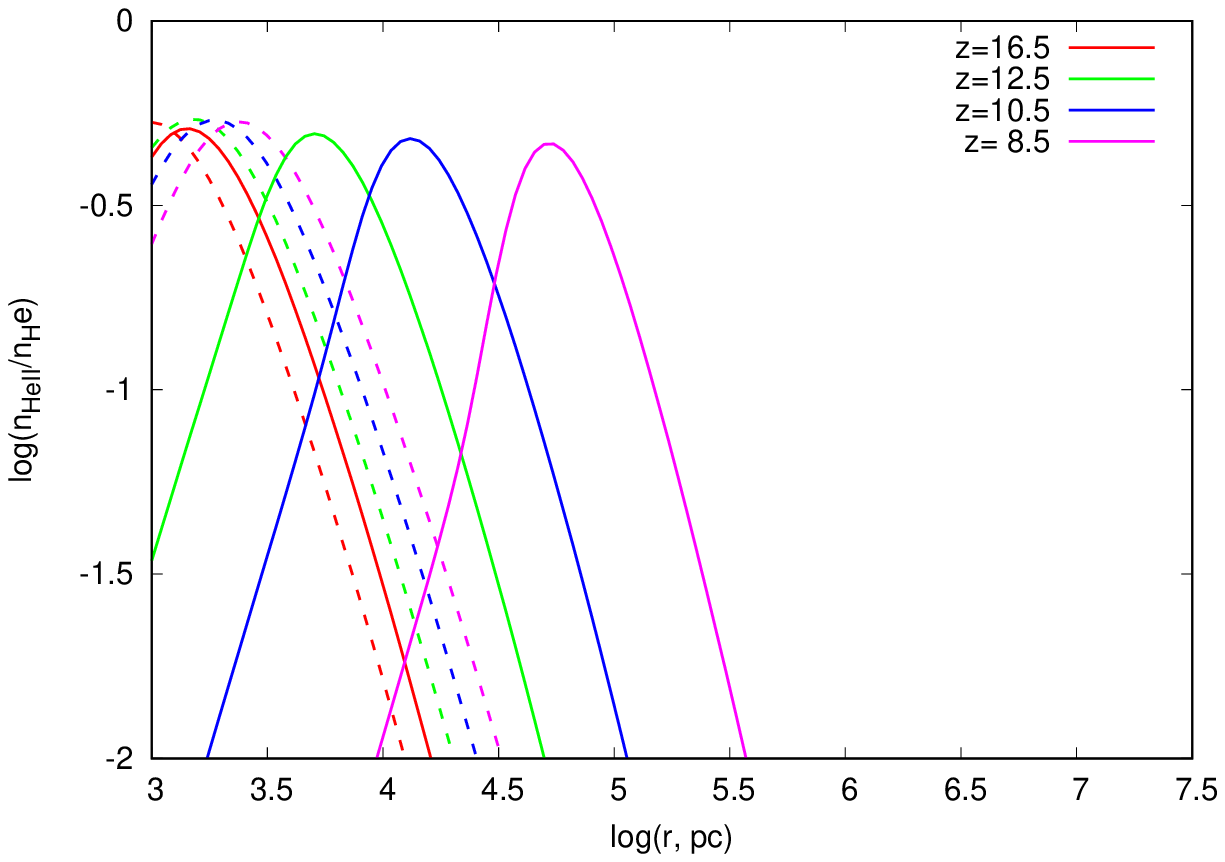}
\caption{
The radial distribution (radius is in physical, not comoving units) of the kinetic temperature (upper), HII fraction (middle), and  HeII fraction (lower) around a BH {with initial mass $M_{BH,z_0}=300\msun$ and radiative efficiency $\epsilon=0.1$} starting its evolution at $z_0 = 20$ for several redshifts: $z = 16.5, 12.5, 10.5, 8.5$ (lines from left to right); {dashed lines stay for a BH with a constant mass {$M_{BH}=300\msun$}.  } 
}
\label{fig-evol}
\end{figure}

Figure~\ref{fig-evol} shows  thermal and ionization evolution around both non-growing (constant mass) and growing BHs with the initial redshift  $z_0 = 20$. The non-growing BH is surrounded by {the zone of influence} -- the region of (physical) size  $r\simlt 10^{5}$~pc, in which  the gas temperature and the ionized fraction of hydrogen and helium  differ significantly  from the background values. {In the central part of $r\sim 10$ kpc} the  ionized fraction of hydrogen reaches close to unity  and the temperature exceeds  $10^4$~K. {The zone of ionized gas shows} a slow evolution with redshift, {$r\propto t$}. 

The growing  BH produces more ionizing photons and  the { zone of influence increases in time much faster than in the previous case, nearly as $r\propto t^{1.6}$}. For $\epsilon=0.1$ the size of a sphere, where the gas temperature and ionizing fraction differ markedly from the background values, becomes at redshift $z=8.5$ more than an order of magnitude larger as compared to that for a non-growing BH. The physical {size of the zone of influence} in the latter case grows from nearly 10~kpc to 300~kpc during the growth of the central BH. The {ionization fraction of hydrogen can reach a few percents within the region with temperature exceeding 300~K}.  The zone of influence of helium is generally smaller and reaches on the upper end 100~kpc. 

\begin{figure}
\center
\includegraphics[width=85mm]{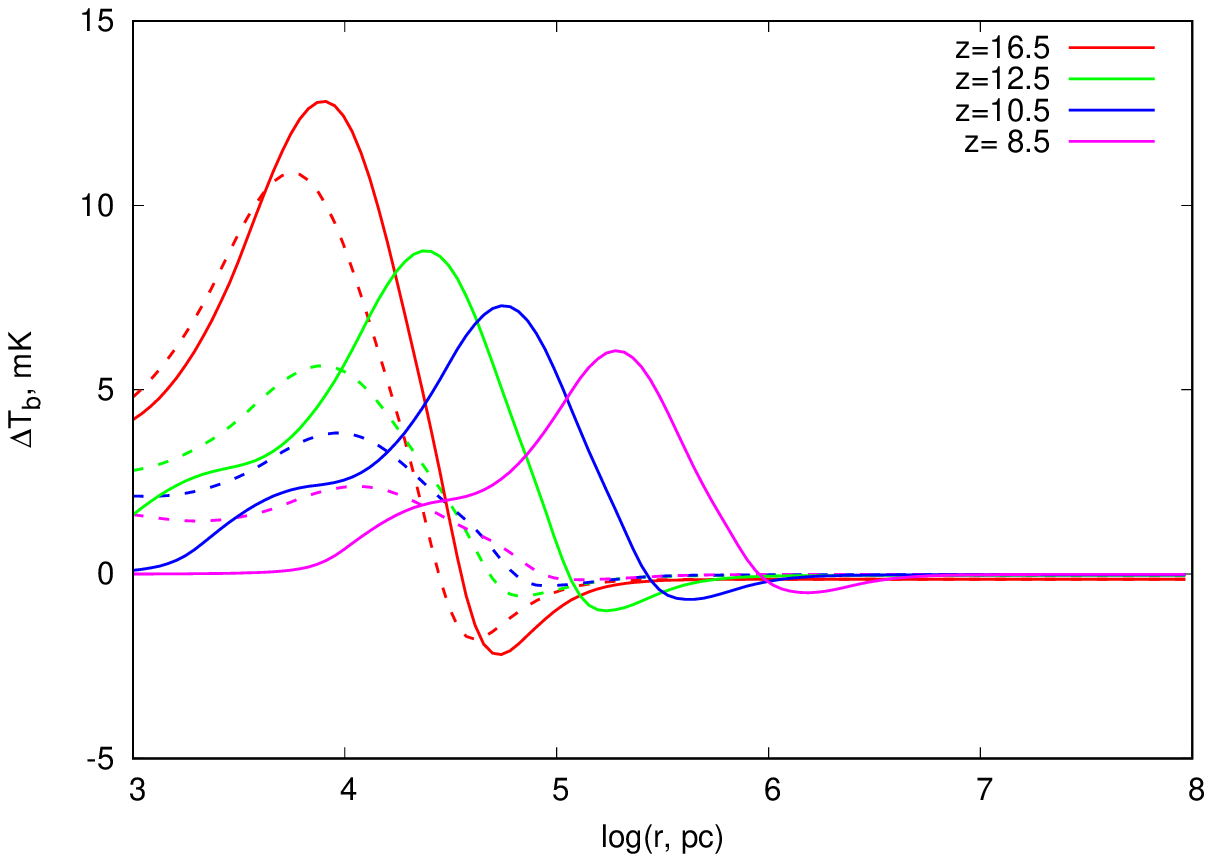}
\includegraphics[width=85mm]{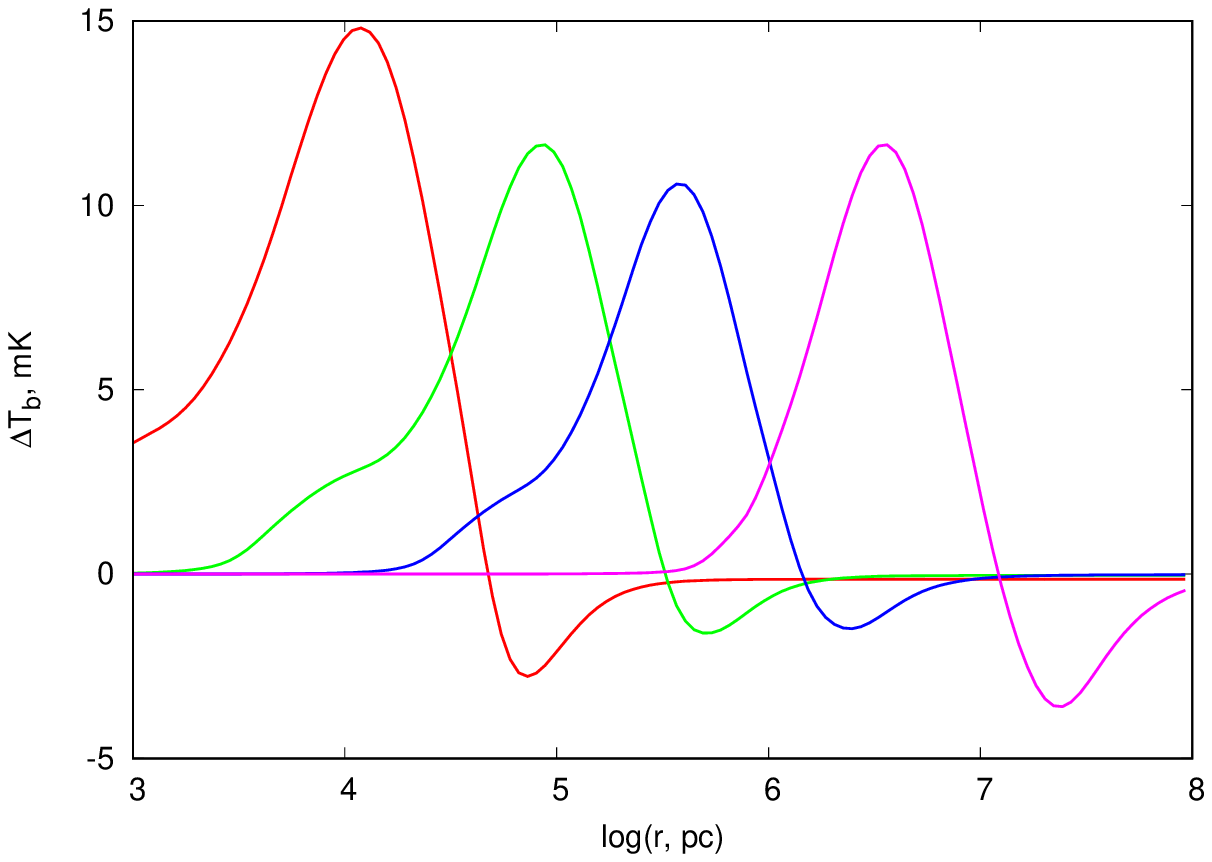}
\caption{
{The brightness temperature in the 21 cm HI line as a function of radius is plotted for different cases around a BH with initial mass $M_{BH,z_0}=300~\msun$, starting its evolution at $z_0 = 20$ with the radiative efficiency $\epsilon = 0.1$ (upper panel) and 0.05 (lower panel).  The dashed lines {shown in the upper panel} correspond to a non-growing BH with constant BH mass $M_{BH}=300~\msun$.  }}
\label{figh1ya}
\end{figure}

First, we consider the case when the implications of EDGES results are not taken into account. Figure~\ref{figh1ya} shows radial profiles of brightness temperature for a static (dashed lines) and growing (solid lines) BHs for two different values of the radiative efficiency $\epsilon=0.1$ and $0.05$. As seen the brightness temperature peaks at the region with sufficiently high kinetic gas temperature $T$ and high fraction of the atomic hydrogen, i.e. where the product $Tx_{\rm HI}$ peaks. On the contrary, the signal from 21 cm vanishes  where  the Lyman-$\alpha$ coupling becomes inefficient. As seen from Eq.~(\ref{epsi}) lower $\epsilon$ causes higher accretion rate and higher luminosity. Consequently, the zones of influence are greater and the 21 cm line emission brightens at a given time, such that its appearance becomes more clearly pronounced.

\begin{figure}
\center
\includegraphics[width=85mm]{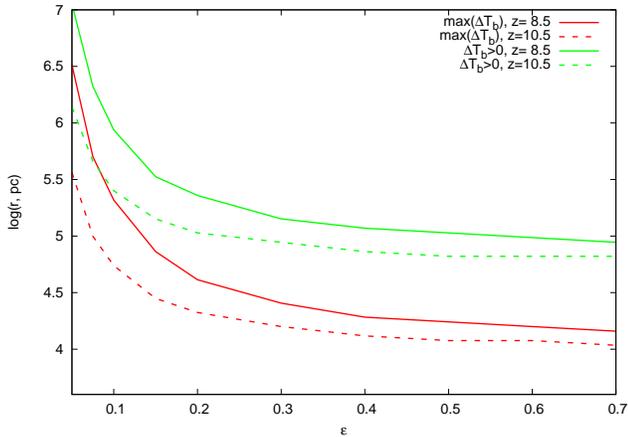}
\caption{
The radii of spheres around a BH {with the initial mass $M_{BH,z_0}=300~\msun$} starting its evolution at $z_0 = 20$ versus the radiative efficiency $\epsilon$ at two redshifts: $z = 10.5$ (dashed lines) and 8.5 (solid lines). The red lines show the radius at which the brightness temperature in the 21 cm HI line $\Delta T_b$ reaches maximum and the green lines depict the radius at which $\Delta T_b$ is positive (see Figure~\ref{figh1ya}). }
\label{fig-rad}
\end{figure}

\begin{figure}
\center
\includegraphics[width=85mm]{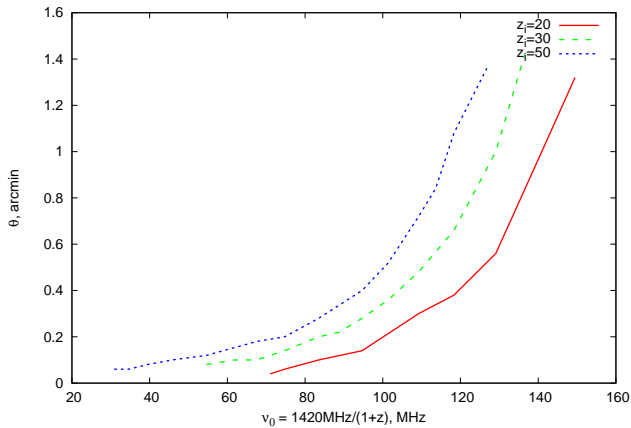}
\caption{
The dependence 'angular diameter -- observed frequency' for spheres emitted in the HI 21 cm line around a growing BH with radiative efficiency $\epsilon = 0.1$ starting its evolution at redshifts $z_0 = 50$, 30 and 20 (lines from left to right). The diameter of spheres is determined, where the brightness temperature in the 21 cm HI line $\Delta T_b^{\rm HI}$ reaches maximum (see Figure~\ref{figh1ya}).
}
\label{fig-adia-fre-HI}
\end{figure}

Figure~\ref{fig-rad} shows the radius at which the brightness temperature in the 21 cm HI line $\Delta T_b$ reaches maximum (red lines) and the radius beyond which $\Delta T_b$ becomes negative (green lines) versus the radiation efficiency $\epsilon$, at two redshifts: $z = 10.5$ (dashed lines) and 8.5 (solid lines) for a BH starting its evolution at $z_0 = 20$. Clearly seen is that the region influenced by growing BHs is larger for smaller $\epsilon$: for $\epsilon\simeq 0.05$ it extends up to $\sim 1$~Mpc, corresponding  to the comoving scale
$\simeq 10 \, \rm Mpc$ which is close to the spatial resolution of on-going radio-interferometers like LOFAR. For $\epsilon \sim 0.1\hbox{--}0.25$ it reaches around  $0.1$~Mpc at $z\simlt 10.5$. This is also comparable  to the mean distance between minihalos with $M\simgt 10^6~\msun$ at $z\sim 10$, what means that a growing BH can affect star-formation in neighbouring minihalos \citep{haiman00}, that can result in a stronger signal in the 21 cm line. 

The evolution of such region can be represented in observable values. Figure~\ref{fig-adia-fre-HI} shows how the angular diameter of the region emitting in 21 cm line depends on observed frequency $\nu_o = 1420~{\rm MHz}/(1+z)$ for a growing BH with radiative efficiency $\epsilon = 0.1$ starting its evolution at redshifts $z_0 = 50$, 30 and 20. The diameter of the emitting sphere is defined as that where the brightness temperature in the 21 cm HI line $\Delta T_b^{\rm HI}$ reaches the maximum (see Figure~\ref{fig-evol}). One can note that the angular size of the regions becomes greater than 1~arcmin at $\nu \sim 110-150$~MHz. The increase of radiative efficiency obviously leads to larger angular size, e.g. it grows up to 1.7~arcmin at 150~MHz ($z=8.5$). 

Ongoing radio interferometer  such as LOFAR and upcoming SKA1-LOW have the capability of detecting the contrast between HI brightness temperature on angular scales of a few arcminutes. { This contrast could be detected statistically, e.g. by measuring two-point correlation function of the intensity of the redshifted HI line, or by  imaging\footnote{for a discussion on sensitivities for these two observables in radio interferometry, see e.g. \citet{2008ApJ...673....1S}}. Our analysis can be extended to predict the two-point functions of the spatial distribution of HI but we do not attempt it here partly because these functions depend on the fraction of the universe in which the thermal and ionization history is  impacted by early BHs  \citep[e.g.][]{Zaldarriaga2004}; this fraction cannot be reliably computed because the space density of the precursor of these  BHs  is highly uncertain, as already  noted above.   For  imaging, the projected  sensitivity of SKA is expected to reach a few millikelvins on angular scales from 1--10 arcminutes \citep{2015aska.confE...1K}}. The expected contrast (Figure~\ref{figh1ya}), particularly in light of the recent EDGES result, is likely  to reach a few hundred milli kelvins which is easily detectable by SKA1-LOW and possibly by LOFAR.


\begin{figure}
\center
\includegraphics[width=85mm]{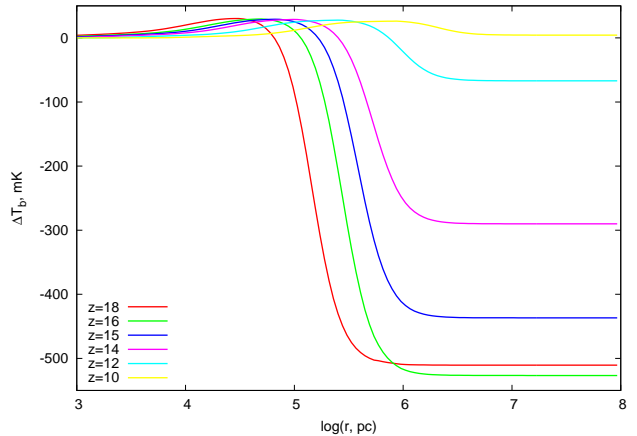}
\caption{The brightness temperature in the 21 cm HI line vs radius around a BH with initial mass $M_{BH,z_0}=300~\msun$, $\epsilon = 0.1$ and $z_0 = 40$, with altered thermodynamics of baryons due to elastic scattering with cold dark matter. }
\label{figh2ya}
\end{figure}

\begin{figure}
\center
\includegraphics[width=85mm]{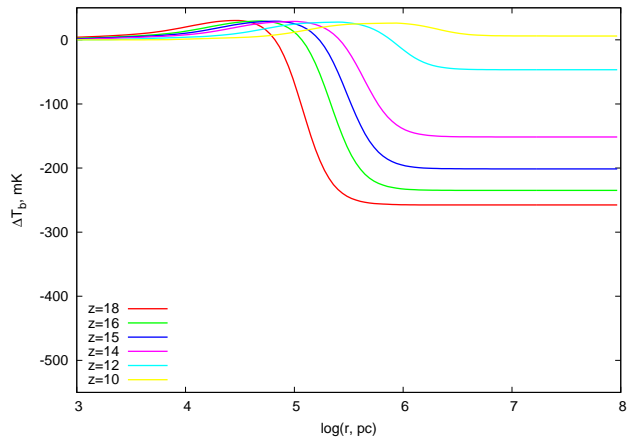}
\caption{
Same as in Figure \ref{figh2ya} without baryon cooling due to elastic collisions with dark matter, but with an additional heating coming from energy released by stars in the first episode of star formation.  
}
\label{fig-rad-vsz040}
\end{figure}

\begin{figure}
\center
\includegraphics[width=85mm]{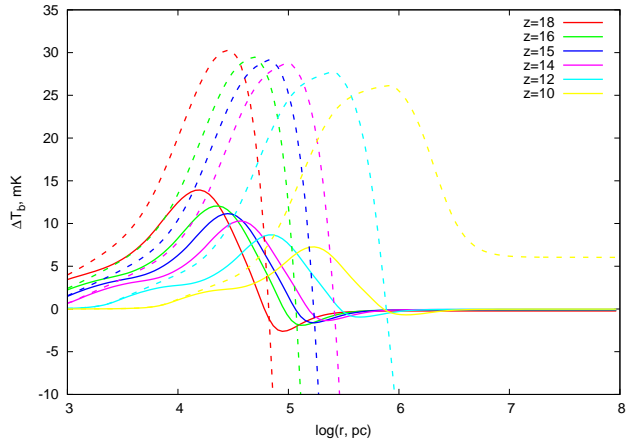}
\caption{
{ The brightness temperature in the 21 cm HI line vs radius around a BH with initial mass $M_{BH,z_0}=300~\msun$, $\epsilon = 0.1$ and $z_0 = 40$, without (solid lines) and with (dashed lines) an additional heating coming from energy released by stars in the first episode of star formation. }
}
\label{fig-rad-vsz040-heatc}
\end{figure}

\subsubsection{Altered thermodynamics from DM cooling} \label{susub:dm} 

We discuss next the impact of altered  thermal history on cosmological observables caused by additional cooling of baryons in elastic interactions with dark matter \citep{2018Natur.555...67B}. We show the impact of global thermal state of the neutral gas implied by this result in Figure~\ref{figh2ya}. Closer to the BH the thermal and ionization state of the gas is determined by the emission from the BH. However, unlike the case shown in Figure~\ref{figh1ya}, the HI signal  is seen in strong absorption far away from the BH in the redshift range 15--19, in  agreement with EDGES results discussed above.   

For comparison in Figure~\ref{fig-rad-vsz040} we show a similar model without baryon cooling forced by elastic interactions with dark-matter, but with heating from energy released in initial episodes of star formation as in \citep{barkanafi,cohenfi}. An obvious distinctive feature of models with DM cooling is that outside the zone of influence the brightness temperature follows the global behavior of HI spin temperature.  This situation causes strong spatial contrast in HI brightness temperature in the redshift range of interest,  which makes it easier to observe this signal using ongoing and future radio interferometers.

\subsection{Dependences on initial parameters} \label{ipar}

Minihalos are thought to  form in high peaks of the cosmological density field. Even though for higher redshifts such peaks become rarer,  minihalos can form as early as  $z\sim 50$ \citep{gao-first05}. Such minihalos can host first BHs, which in turn can become progenitors of supermassive BHs $M\sim 10^9~\msun$ found at $z\sim 6-7$ \citep[e.g.][]{mortlock11,wu-bhs15}. {We briefly discuss possible observational manifestations from BHs began growing at higher redshifts.}

One obvious consequence of a BH growing at higher redshift is the larger radius of the zone of influence at a given $z$. For instance, the size of the zone around a BH growing from $z_0=25$ is greater than that of a BH  at  $z_0=20$ by about 60\% {at $z=16.5$ (the corresponding 21 cm line shifts to 80~MHz) and 30\%  at $z=9$ (the corresponding line peak frequency is 142~MHz).} The brightness temperature magnitude decreases from 5.7~mK at 80~MHz to 1.8~mK at 142~MHz, with  weak dependence on  the initial redshift $z_0$. 

Another important issue concerns the mass budget of a growing BH. Dark matter halos that host BHs should  have sufficient amount of baryons to feed the BH. This requirement is especially critical for {lower values} of radiative efficiency $\epsilon$. As  mentioned above, for $\epsilon=0.1$, a BH mass  grows by roughly a factor  $2.5\times 10^3$ in $\sim$400~Myr evolution, but  the factor reaches  $1.5\times 10^7$ for $\epsilon=0.05$. In such halos the HI column density might be higher than the fiducial value adopted here,  $N_{\rm HI} = 10^{20}$~cm$^{-2}$. {An increase of the HI column density makes the brightness temperature radial profiles to shrink. For example, for $N_{\rm HI} = 10^{21}$~cm$^{-2}$},  the peak of 21 cm brightness temperature for $z=16.5$ shifts from $r\sim 7$~kpc, corresponding to the fiducial HI column density, to $\sim 3$~kpc. However, later on this difference diminishes, e.g. the ratio between the radii becomes about 1.3 at $z=8.5$. The difference in sizes of the zone where $\Delta T_b>0$, is smaller and becomes negligible for the final redshift. 

The masses of stellar BHs formed by very massive stars remains uncertain. It is conceivable that the initial mass of a BH may be { either lower or} higher than the fiducial value  $M_{BH,t=0} = 300~\msun$. As expected more massive BHs produce larger zones of influence. { The radius of the zone at which the brightness temperature in HI 21 cm line reaches maximum depends on the initial mass of a BH seed as $r \sim M_{BH,t=0}^{0.38}{\rm exp}(z^{-1.16})$ for $\epsilon=0.1$ and $M_{BH,t=0}=(30-10^3)~\msun$. For example, } its radius increases by  about a factor 1.5 for $M_{BH,t=0} = 10^3~\msun$ until $z=8.5$ if a BH starts growing $z_0=20$. 

The size of the zone of influence around a growing BH depends on the slope $\alpha$ of the spectral energy distribution (\ref{lum}), which might vary from $-1.7$ to $-1.4$. A flatter spectrum leads to a larger radius of the zone, whereas a steeper one produces a smaller zone. For instance, the zone around a growing BH with $\alpha=-1.7$ is $\sim 20-25$\% smaller than that for the fiducial value $\alpha=-1.5$.

Finally, we consider how heating and Ly$\alpha$ background affect evolution of zones of influence around growing BHs. Resonance and high-energy photons produced due to the initial episode of star formation provide a homogeneous background in this case. Figures~\ref{figh1ya} and \ref{fig-rad-vsz040} show the HI signal around a halo with a growing BH immersed in the IGM evolved adiabatically and exposed to both X-ray and Ly$\alpha$ background photons as in \citet{barkanafi,cohenfi}. These models represent  extreme cases: in the former there is no external background radiation, whereas the latter includes strong (maximum in the sense that $T\simeq T_s$) Ly$\alpha$ pumping rate and heating from background ionizing photons. We combine the expected HI signals for these models in Figure~\ref{fig-rad-vsz040-heatc}. The growth of the brightness temperature in the model with heating is due to strong Ly$\alpha$ background. The size of the influence zone increases with decreasing redshift in presence of the background. At $z=10$ the size doubles as compared to that in the model without the background radiation. At high redshifts, where heating is weak such an increase is small.

\subsection{$^3${\rm HeII} hyperfine line} \label{sub:3he}

\begin{figure}
\center
\includegraphics[width=85mm]{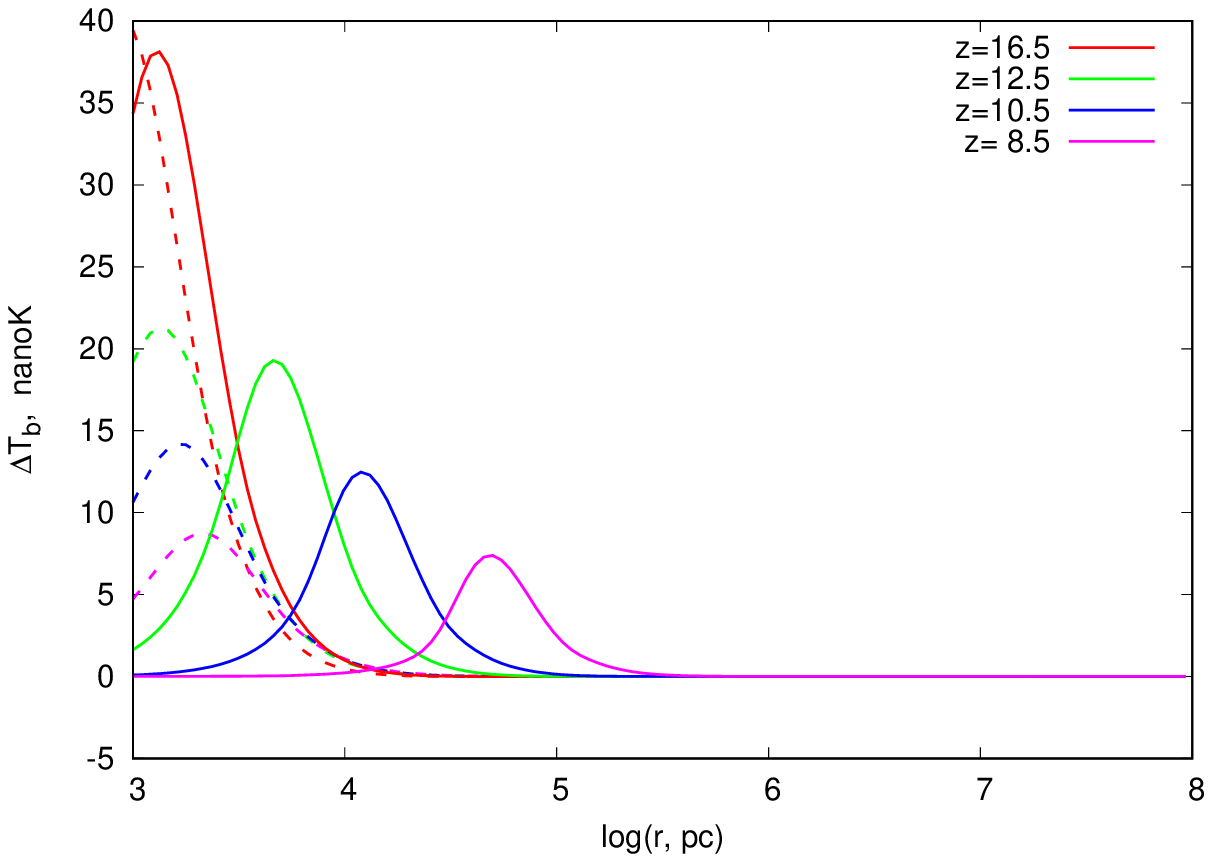}
\includegraphics[width=85mm]{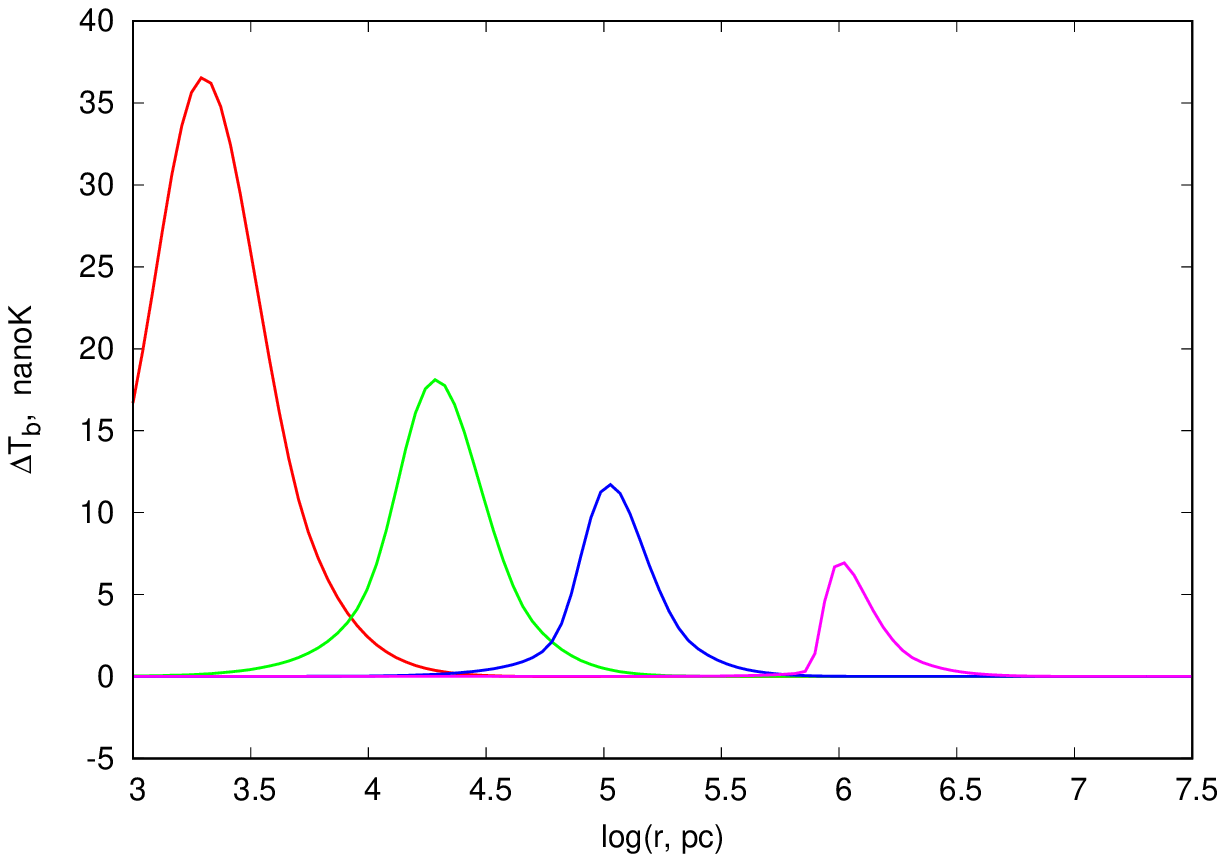}
\caption{{Same as in \ref{figh1ya} for $^3$HeII 3.46~cm line.}}
\label{figh1he3}
\end{figure}

As discussed above the other potential observable is the $^3$HeII hyperfine line. 
Unlike massive stars, { which can also ionize HeII \citep[e.g.,][]{tum00}}, BHs can form {large} HeII and even HeIII ionization zones.  Figure~\ref{fig-evol} presents the radial distribution of the HeII fraction around BHs with a constant and a growing mass. The size of the HeII region around BH with constant mass of several hundreds solar masses is about 1-3~kpc, that is compared to the virial radius of the host dark matter minihalo. However, it increases by several ten or even hundred times around a growing BH. Such zones can emit in the hyperfine structure line of a singly ionized helium-3 isotope. The brightness temperature in the $^3$HeII line reaches several tens nanoK (Figure~\ref{figh1he3}),  and the size of the emission zone can extend up to more than 10~kpc. 

\begin{figure}
\center
\includegraphics[width=85mm]{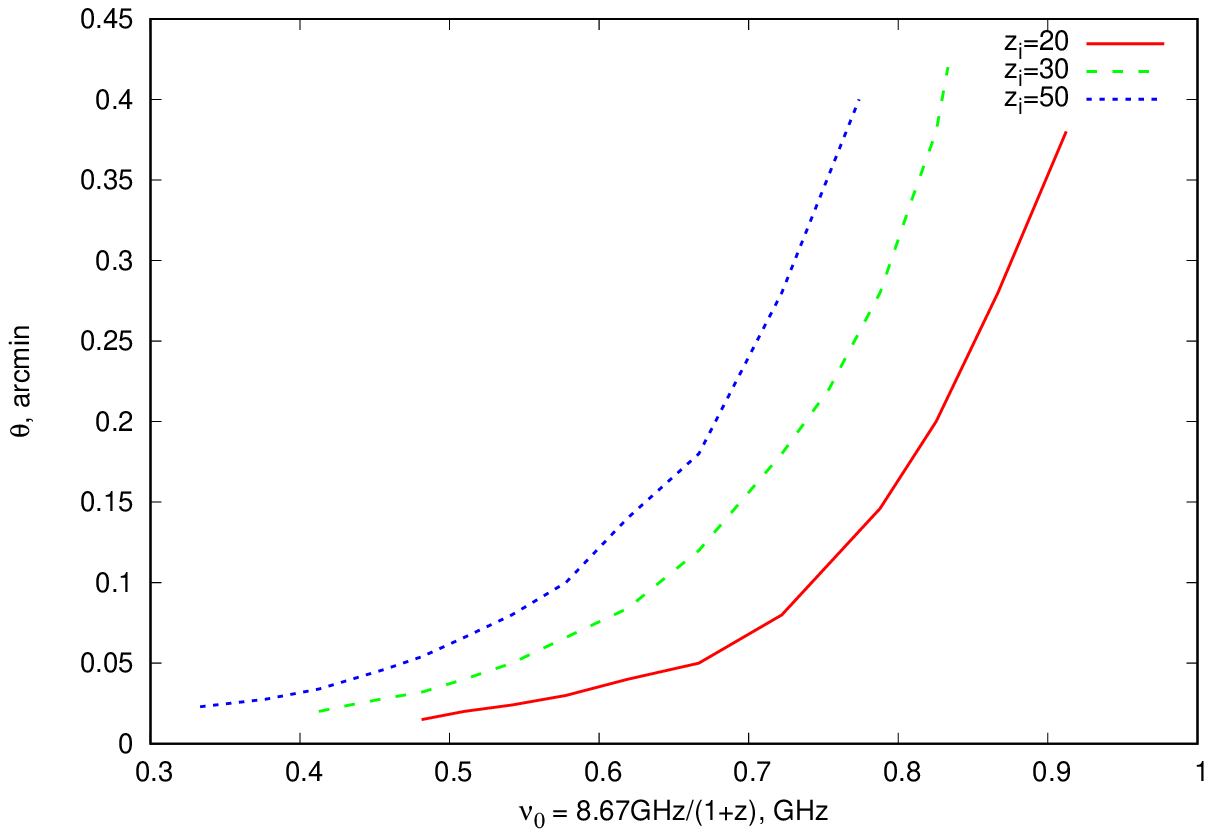}
\caption{
The dependence 'angular diameter -- observed frequency' for spheres emitted in the $^3$HeII 3 cm line around a growing BH with radiative efficiency $\epsilon = 0.1$ starting its evolution at redshifts $z_0 = 50$, 30 and 20 (lines from left to right). The diameter of spheres is defined as where the brightness temperature in the line $\Delta T_b^{\rm 3HeII}$ reaches maximum (see Figure~\ref{figh1he3}).
}
\label{fig-adia-fre-3HeII}
\end{figure}

The angular size of such zones at frequency $\sim 1$ GHz is of $0.3\hbox{--}0.4$ arcmin as shown in Figure~\ref{fig-adia-fre-3HeII}. Upcoming radio interferometer SKA1-MID can reach flux sensitivity of sub micro-Jansky at such frequencies  at these angular scales \citep{2015aska.confE...3A}, which corresponds to brightness temperature sensitivity which is still orders of 
magnitude larger than the expected signal. Therefore it is unlikely this signal would be detected by  upcoming radio interferometers.   

\begin{figure*}
\center
\includegraphics[width=85mm]{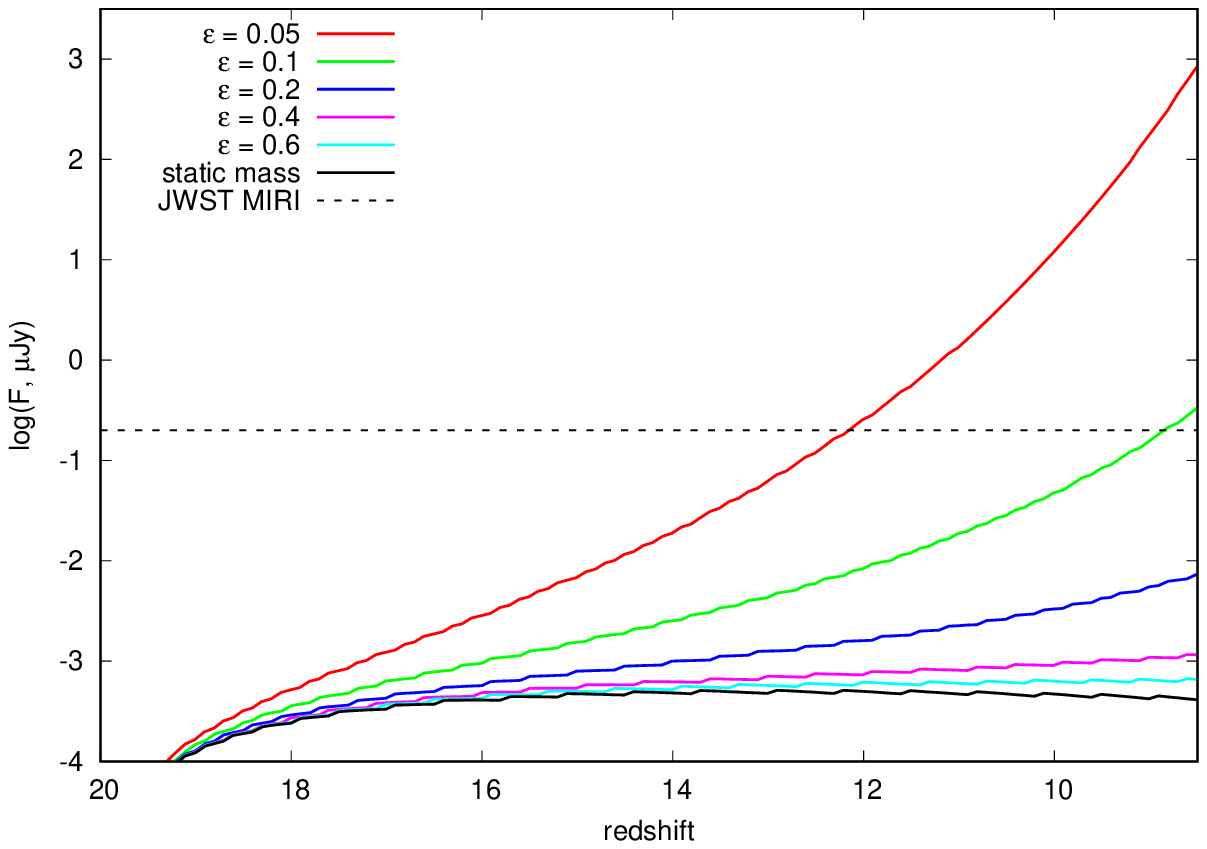}
\includegraphics[width=85mm]{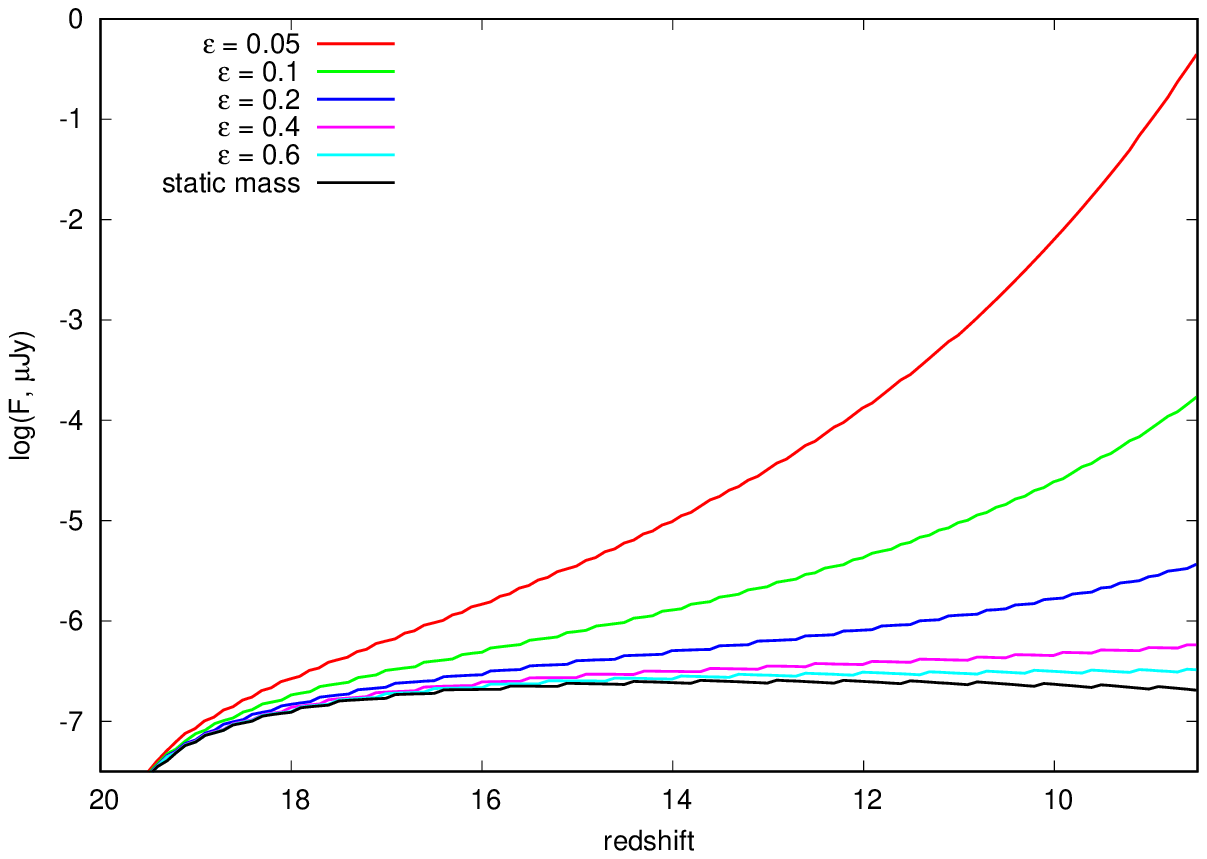}
\caption{
{The fluxes in H$\alpha$ (left panel) and Hn$\alpha, n=30$ (right panel) recombination lines that can be detected from partially ionized spheres around a BH {with the initial mass $M_{BH,z_0}=300~\msun$} starting its evolution at $z_0 = 20$ for several radiative efficiency $\epsilon = 0.05, 0.1, 0.2, 0.4, 0.6$ and static BH mass {$M_{BH}=300~\msun$} (lines from top to bottom)}.
}
\label{fig-flx}
\end{figure*}

\subsection{$n\alpha$ {\rm HI} recombination lines}

We next consider recombination lines arising from ionized regions surrounding the accreting BH. 

Figure~\ref{fig-flx} presents  flux in H$\alpha$ line (left panel) from partially ionized spheres around a BH starting its evolution at $z_0 = 20$. The size of regions that dominate emission in H$\alpha$ line is $\simeq 10 \, \rm kpc$.  The flux for  $\epsilon \sim 0.05$ exceeds  $\mu$Jy at $z\simlt 11$. The angular size of such a region is $\simeq 0.2''$ which can be resolved by JWST, which has angular resolution of $\simeq 0.05''$. The redshifted H$\alpha$ line has wavelength $\simeq 7.8$~micron for this case which is accessible  to mid-Infrared  (MIRI) instrument aboard  JWST. Around  this wavelength,  a source of flux  $\simeq  0.2 \, \rm \mu$Jy can be detected with signal-to-noise $S/N=10$ in integration time of $10^4 \, \rm sec$. As this sensitivity corresponds to source  within the resolution element of the instrument ($\simeq 0.05''$) and the source angular size is nearly four times the resolution, the sensitivity of detection degrades  by  nearly a factor of 4. A comparison of this estimate of sensitivity with fluxes shown in Figure~\ref{fig-flx} shows that  the regions of influence around BHs with $\epsilon \sim 0.05$ (starting its growth at $z_0=20$) can be detected for  $z\simeq 10\hbox{--} 12$.  For BHs with higher radiation efficiency $\epsilon \sim 0.1$ the surrounding gas might be   observable  at $z\sim 8.5$ in H$\alpha$.

Expected fluxes from transitions for $n=30$ (Figure~\ref{fig-flx}, right panel) are near the thresholds of modern radio telescopes only around very rapidly growing BHs ($\epsilon=0.05$). It should be noted that it is much easier to detect a $n,n-1$ transition for smaller  $n$ as the flux of the line  $\varepsilon \sim n^{-2.72}$. 

\section{Conclusions}  \label{sec:sumcon}

In this paper, we considered the impact of a {growing} black hole on thermal and ionization state of the IGM in the redshift range $8 < z <25$, and discuss possible observables that can probe this influence. We have found that the sizes of {zones of ionized gas} around growing BHs are greater as compared to that for a non-growing BH: for accretion with radiative efficiency $\epsilon=0.1$ they are more than order of magnitude larger at redshift $z=8.5$. The physical size of a zone of influence increases from nearly 10~kpc to 300~kpc during the growth of a BH. The most part of this region contains highly ionized hydrogen upto a reasonable fraction of unity, and temperature exceeding 300~K. Helium ionization region is generally smaller and reaches a maximum of 100~kpc. 

We consider three observables as probe of growing primordial BHs. 

We show that the influence region of 21~cm emission around an accreting  BH with  radiative efficiency  $\epsilon\simgt 0.05\hbox{--}0.1$ could be in the range of a few hundred kilo-parsecs to 1 Mpc (Figure~\ref{figh1ya}). The angular scale of this emission and the spatial contrast of the HI signal is accessible to ongoing and upcoming radio telescopes such as SKA1-LOW. We also consider the impact of recent EDGES observation  \citep{2018Natur.555...67B} and show that it greatly enhances the expected contrast (Figure~\ref{figh2ya}). 

We also study  the emission of  hyperfine line of  $^3$HeII ($\lambda = 3.4 \, \rm cm$) from regions surrounding the growing BH. The brightness temperatures in these lines could reach tens of  nano-Kelvin. Taking into account the sizes of these regions we anticipate that this emission cannot be detected by upcoming radio telescopes SKA1-MED. 

We finally consider hydrogen recombination lines (n,n-1) from ionized regions surrounding growing BHs. The H$\alpha$ line provides the best prospect of detection (Figure~\ref{fig-flx}); JWST can detect this line with $S/N=10$ in ten thousand seconds of integration. Expected fluxes from transitions between higher levels (e.g. Figure~\ref{fig-flx} for $n= 30$)  are near the thresholds of modern radio telescopes only around very rapidly growing BHs. 

In sum: we model emission from an accreting primordial BH and study its impact on the ionization and thermal state of surrounding  medium. We also consider the prospects of the detection of this dynamical process in the redshift range $8.5 < z < 25$.

In conclusion we note that the observability of the features we discuss in the paper would be greatly boosted  if the precursors of supermassive black holes could be detected at high redshifts. This possibility has been studied by  \citet{valiante18sta,valiante18obs}. Their analysis suggests that future missions such as JWST will be able to detect high-mass BH seeds at $z\sim 16$ directly.

\vspace{1cm}

We are thankful to the referee for a careful reading of the manuscript and very detailed comments.
This work is supported by the joint RFBR-DST project (RFBR 17-52-45063, DST P-276). The work by YS is done under partial support from the joint RFBR-DST project (17-52-45053), and the Program of the Presidium of RAS (project code 28). The code for the thermal evolution has been developed under support by Russian Scientific Foundation (14-50-00043).


\end{document}